\documentclass{amsart}

\allowdisplaybreaks[1]
\numberwithin{equation}{section}
\newcommand{\Li}[1]{{\rm Li}_{#1}}   
\newcommand{\Cl}[1]{{\rm Cl}_{#1}}
\newcommand{\Gl}[1]{{\rm Gl}_{#1}}
\newcommand\half{\tfrac{1}{2}}
\newcommand{\abs}[1]{\lvert#1\rvert}
\newcommand{\seriesref}[1]{Series~\ref{#1}}
\newcommand{\secref}[1]{\S\ref{#1}}
\newcommand{\appref}[1]{Appendix~\ref{#1}}
\renewcommand{\Re}{{\rm Re}}
\renewcommand{\Im}{{\rm Im}}

\newtheorem*{theorem}{Theorem}
\newtheorem*{corollary}{Corollary}
\newtheorem*{lemma}{Lemma}
\newtheorem{series}{Series}

\begin{document}
\begin{flushright}
{DESY 97-245}\\
{hep-th/9801168}\\
{December 1997} \\
\end{flushright}
\vspace*{1cm}
\title[One- and two-dimensional Series]{Summing one- and 
two-dimensional Series related to the Euler Series}

\author{Odd Magne Ogreid}
\address{Department of Physics \\  University of Bergen, All\'egt.~55 \\
N-5007 Bergen\\ Norway}
\email{Odd.Ogreid@fi.uib.no}
\urladdr{http://www.fi.uib.no/\~{}ogreid/}
\author{Per Osland}
\address{Department of Physics \\  University of Bergen, All\'egt.~55 \\
N-5007 Bergen\\ Norway}
\curraddr{Deutsches Elektronen-Synchrotron DESY \\ D-22603 Hamburg, Germany}
\email{Per.Osland@fi.uib.no}
\urladdr{http://www.fi.uib.no/\~{}osland/particle.html}

\thanks{This research has been supported by the Research Council of Norway,
and by DESY}
\begin{abstract}
We present results for some infinite series appearing in Feynman diagram 
calculations, many of which are similar to the Euler series.
These include both one-dimensional and two-dimensional series.
Most of these series can be expressed in terms of 
$\zeta(2)$, $\zeta(3)$, the Catalan constant $G$ and $\Cl{2}(\pi/3)$ where
$\Cl{2}(\theta)$ is Clausen's function. 
\end{abstract}
\maketitle

\section{Introduction}
\label{s:intro}
When calculating radiative corrections in Quantum Field Theory, one will 
encounter multi-dimensional Feynman integrals \cite{Feynman}. 
These often present considerable mathematical challenges.
Several methods are available for such calculations.
Unfortunately, the evaluation of such integrals is a tedious task 
and a simple result cannot always be found. 

A powerful method for doing Feynman integrals consists in using Mellin 
transforms \cite{Myint,Oberhettinger}. This approach is particularly 
useful when one wants to expand the result in powers of logarithms 
of the kinematical variables.
One may thus factorize the integrands to be left with a number of complex
contour integrals. These can in turn be evaluated by means of 
residue calculus. Upon calculating these contour integrals,  
the result will be expressed as infinite series over one or more summation 
variables. For typical applications to Quantum Field Theory, see
e.~g.\ \cite{ChengWu}. 

As a result of Feynman integrals, the Riemann zeta function
\cite{Abramowitz,Ahlfors}:
\begin{equation*}
\zeta(z)=\sum_{n=1}^\infty\frac{1}{n^z}
\end{equation*}
appears frequently.
In particular, we shall need its values for $2$ and $3$:
\begin{align*}
&\sum_{n=1}^\infty\frac{1}{n^2}=\zeta(2)=\frac{\pi^2}{6}
=1.644\; 934 \dots,\\
&\sum_{n=1}^\infty\frac{1}{n^3}=\zeta(3)
=1.202\; 057 \dots.
\end{align*} 
In Feynman integrals, these constants often appear as results of 
integrations where polylogarithms are involved.

The above mentioned method, using Mellin transforms, has been applied 
to Feynman integrals appearing in
two-loop studies of Bhabha scattering \cite{Bhabha,Bhabha-two-loop}.
In recent work \cite{Ogreid}, the transcendental constants appear when 
summing series instead of resulting from integrations. 

A large number of the series encountered are not found in the familiar 
tables. This is the 
case for some of the one-dimensional series, and in particular for the 
two-dimensional ones. The purpose of this article is to present results 
for some of the series encountered. 
These can be expressed in terms of a few  
constants, including $\zeta(2)$, $\zeta(3)$ and the Catalan constant $G$. 
In addition, the constant $\Cl{2}(\pi/3)$ appears frequently. 
It seems to us that few of these results are known, 
in particular none of those involving Clausen's function. 
Some basic properties of Clausen's function are given in 
\appref{a:specfunc}.
Similar results, including some of those given here, have been presented in 
\cite{Berndt,Borwein,Borwein-etal,Doelder,Shen,Sita}. An excellent article
on triple Euler series can be found in \cite{Borwein-preprint}. Here, 
the reader will also find an appendix written by D.~Broadhurst
on the connection between Euler series,  Quantum Field Theory and Knot Theory.
For further references on such series, we refer the reader to \cite{Hoffman}.

Before we turn to the evaluation of these sums, we note that in this
work we will interchange integrations, differentiations, sums
and limits at will. In general, care should be exercised in doing so.
For the calculations shown here, all these interchanges are allowed.

\section{One-dimensional series}
\label{s:one-d}
Let us consider the family of series 
\begin{equation*}
\sum_{n=1}^\infty\frac{1}{n^2}[\gamma+\psi(1+kn)]
=\sum_{n=1}^\infty\sum_{j=1}^{kn}\frac{1}{n^2j},
\end{equation*}
where $\gamma$ is the Euler constant,
$k$ is a positive integer and $\psi(z)$ is the logarithmic derivative 
of the gamma function [see \appref{a:specfunc}]. 
For $k=1$, this reduces to one of the double Euler series.
Our family of series can all be expressed in terms of a rational multiple
of $\zeta(3)$ and a finite sum over Clausen's function \cite{Lewin}.
This is true also for the corresponding alternating series:
\begin{theorem}
\begin{align}
\sum_{n=1}^\infty\frac{1}{n^2}[\gamma+\psi(1+kn)]
&=\left(\frac{k^2}{2}+\frac{3}{2k}\right)\zeta(3)
+\pi\sum_{j=1}^{k-1}j\, \Cl{2}\left(\frac{2\pi j}{k}\right),
\label{e:singlepsisum}
\\
\sum_{n=1}^\infty\frac{(-1)^n}{n^2}[\gamma+\psi(1+kn)]
&=\left(\frac{k^2}{2}-\frac{9}{8k}\right)\zeta(3)
+\pi\sum_{j=1}^{k-1}j\, \Cl{2}\left(\frac{2\pi j}{k}+\frac{\pi}{k}\right), 
\label{e:singlepsisumalternating}
\end{align}
for $k=1,2,3,\dotsc$, where the sums over $j$ are understood to be zero 
when $k=1$.
\end{theorem}

For low values of $k$, the sums over Clausen's function can
be expressed in terms of $\Cl{2}(\pi/3)$ and the Catalan constant
$G$. Explicit results are given in \appref{a:results}.
\begin{proof}
We start with the proof of \eqref{e:singlepsisum}. 
By using the integral representation \eqref{e:psiintegral} 
of the psi function, we can rewrite the series as
\begin{equation*}
\sum_{n=1}^\infty\frac{1}{n^2}[\gamma+\psi(1+kn)]
=\sum_{n=1}^\infty\frac{1}{n^2}\int_0^1{\rm d}t\, \frac{1-t^{kn}}{1-t}
=\int_0^1{\rm d}t\, \frac{\zeta(2)-\Li{2}(t^k)}{1-t},
\end{equation*}  
where we have interchanged integration and summation, thus enabling us 
to express the sum as an integral by using \eqref{e:polylogsum}. 
By the factorization formula \eqref{e:polylogfactorization}, 
the argument of the dilogarithm can be linearized, yielding
\begin{equation*}
\int_0^1{\rm d}t\, \frac{\zeta(2)-k\sum_{j=1}^k\Li{2}(\omega^jt)}{1-t},
\end{equation*}
where $\omega=e^{2\pi i/k}$. If one tries to calculate this integral 
term by term, one will see that each term is divergent although 
the sum converges. For the purpose of splitting up the integral,
we introduce a regulator as follows,
\begin{equation*}
\begin{split}
\int_0^1{\rm d}t\, \frac{\zeta(2)-k\sum_{j=1}^k\Li{2}(\omega^jt)}{1-t}
&=\lim_{x\to 1^-} 
\int_0^1{\rm d}t\, \frac{\zeta(2)-k\sum_{j=1}^k\Li{2}(\omega^jt)}{1-xt}\\
&=\lim_{x\to 1^-}\left[\int_0^1{\rm d}t\, \frac{\zeta(2)}{1-xt}
-k\sum_{j=1}^k\int_0^1{\rm d}t\, \frac{\Li{2}(\omega^jt)}{1-xt}\right].
\end{split}
\end{equation*}
The first of these integrals is trivial, while for 
those under the sum we will need the result for
\begin{equation}
I_2(x,a)=\int_0^1{\rm d}t\, \frac{\Li{2}(a t)}{1-xt}
\label{e:I-integral}
\end{equation}
in the limit $x\to 1^-$. This integral is studied in 
\appref{a:integral}. 
Using the result from eq.~\eqref{e:Iintegral}, we get
\begin{equation*}
\lim_{x\to 1^-}\left[-\frac{\log(1-x)}{x}\zeta(2)+k\sum_{j=1}^k
\left\{\frac{\log(1-x)}{x}\Li{2}(\omega^j)+S_{1,2}(\omega^j)
+\Li{3}(\omega^j)\right\}\right].
\end{equation*}
By using the factorization formula \eqref{e:polylogfactorization},
we get
\begin{equation*}
\begin{split}
&\lim_{x\to 1^-}\left[-\frac{\log(1-x)}{x}\zeta(2)
+\frac{\log(1-x)}{x}\Li{2}(1)+k\sum_{j=1}^kS_{1,2}(\omega^j)
+\frac{1}{k}\Li{3}(1)\right]\\
&=\frac{1}{k}\zeta(3)+k\sum_{j=1}^kS_{1,2}(\omega^j),
\end{split}
\end{equation*}
where the divergent parts cancel and we may let $x\to 1^-$. 
To proceed further, we note that 
$\sum_{j=1}^kS_{1,2}(\omega^j)=
\sum_{j=1}^kS_{1,2}(\omega^{-j})$, since an inversion of the argument 
simply corresponds to performing the sum in reverse order. 
Thus,
\begin{equation*}
\begin{split}
&\frac{1}{k}\zeta(3)+k\sum_{j=1}^kS_{1,2}(\omega^j) 
=\frac{1}{k}\zeta(3)+\frac{k}{2}\sum_{j=1}^k
\left[S_{1,2}(\omega^j)+S_{1,2}\left(\frac{1}{\omega^j}\right)\right]\\
&=\frac{1}{k}\zeta(3)+\frac{k}{2}\sum_{j=1}^k
\left[\Li{3}(\omega^j)-\frac{1}{6}\log^3(-\omega^j)
-\log(-\omega^j)\Li{2}(\omega^j)+\zeta(3)\right],
\end{split}
\end{equation*}
where we have used the identity \eqref{e:S12identity}. 
Next, we may use the factorization formula \eqref{e:polylogfactorization} 
and the fact that $\log(-\omega^j)=i\pi(2j/k-1)$.
This enables us to follow our convention that $\log(-1)=i\pi$.
Thus, we get 
\begin{equation*}
\left(\frac{k^2}{2}+\frac{3}{2k}\right)\zeta(3)+\frac{k}{2}\sum_{j=1}^k
\left[\frac{i\pi^3}{6}\left(\frac{2j}{k}-1\right)^3
-i\pi\left(\frac{2j}{k}-1\right)\Li{2}(\omega^j)\right].
\end{equation*}
The result must be real. Thus, we may drop all 
the imaginary parts which eventually will cancel, and we are left with
\begin{equation*}
\begin{split}
&\left(\frac{k^2}{2}+\frac{3}{2k}\right)\zeta(3)
+\pi\sum_{j=1}^kj\, \Im \left\{\Li{2}(\omega^j)\right\}\\
&=\left(\frac{k^2}{2}+\frac{3}{2k}\right)\zeta(3)
+\pi\sum_{j=1}^kj\, \Cl{2}\left(\frac{2\pi j}{k}\right)\\
&=\left(\frac{k^2}{2}+\frac{3}{2k}\right)\zeta(3)
+\pi\sum_{j=1}^{k-1}j\, \Cl{2}\left(\frac{2\pi j}{k}\right).
\end{split}
\end{equation*}
In the first step we have used the fact that on the unit circle,
the imaginary part of the dilogarithm is Clausen's function, 
which vanishes when the argument is an integer multiple of $\pi$.
This completes the proof of the first part of the theorem.

For the alternating series \eqref{e:singlepsisumalternating} we will 
use a similar procedure. By using the integral representation 
\eqref{e:psiintegral} for the psi function and performing the sum over $n$, 
we are left with
\begin{equation*}
\int_0^1{\rm d}t\frac{\Li{2}(-1)-\Li{2}[(\phi t)^k]}{1-t},
\end{equation*}
where $\phi^k=-1$.
We now follow the same procedure as in the proof of 
\eqref{e:singlepsisum}, except that we let the sum in the factorization 
formula run from $0$ to $k-1$. Hence, we arrive at 
\begin{equation*}
\frac{1}{k}\Li{3}(-1)+k\sum_{j=0}^{k-1}S_{1,2}(\phi\omega^j).
\end{equation*}
Performing the sum in reverse order simply corresponds to the 
substitution $\omega^j\to\omega^{-j}$. Combining this with the fact that 
we could equally well have introduced $\phi^{-1}$ instead of $\phi$, 
we find that we will get
\begin{equation}
-\frac{3}{4k}\zeta(3)+\frac{k}{2}\sum_{j=0}^{k-1}
\left[S_{1,2}(\phi\omega^j)+S_{1,2}\left(\frac{1}{\phi\omega^j}\right)\right]. 
\end{equation}
We use the fact that $\log(-\phi\omega^j)=i\pi[(2j+1)/k-1]$, 
which preserves the convention $\log(-1)=i\pi$, to get
\begin{equation*}
\left(\frac{k^2}{2}-\frac{9}{8k}\right)\zeta(3)+\frac{k}{2}\sum_{j=0}^{k-1}
\left[\frac{i\pi^3}{6}\left(\frac{2j+1}{k}-1\right)^3
-i\pi\left(\frac{2j+1}{k}-1\right)\Li{2}(\phi\omega^j)\right].
\end{equation*}
Again, we know that the result must be real, and we drop all 
the imaginary parts. Thus,
\begin{equation*}
\left(\frac{k^2}{2}-\frac{9}{8k}\right)\zeta(3)
+\pi\sum_{j=1}^{k-1}j\, \Cl{2}\left(\frac{2\pi j}{k}+\frac{\pi}{k}\right).
\end{equation*}
This completes the proof of the second part of the theorem.
\end{proof}
An immediate corollary of this theorem is:
\begin{corollary}
\begin{align}
\sum_{n=1}^\infty\frac{1}{n^2}[\gamma+\psi(kn)]
&=\left(\frac{k^2}{2}+\frac{1}{2k}\right)\zeta(3)
+\pi\sum_{j=1}^{k-1}j\, \Cl{2}\left(\frac{2\pi j}{k}\right),\\
\sum_{n=1}^\infty\frac{(-1)^n}{n^2}[\gamma+\psi(kn)]
&=\left(\frac{k^2}{2}-\frac{3}{8k}\right)\zeta(3)
+\pi\sum_{j=1}^{k-1}j\, \Cl{2}\left(\frac{2\pi j}{k}+\frac{\pi}{k}\right),
\end{align}
for $k=1,2,3,\dotsc$, where the sum over $j$ vanishes for $k=1$.
\end{corollary}
\begin{proof}
This corollary follows immediately from the theorem by using the 
recurrence formula \eqref{e:psirecurrence} for the psi function.
\end{proof}

For certain low values of $k$, the sums over Clausen's function
may be simplified. Thus, we may state exact and compact results 
for a considerable number of series. Such results are collected in 
\appref{a:results}.

In the Theorem, $n^2$ appears in the denominator of the summand. For higher
powers of $n$, part of the same procedure
can be carried out when summing the corresponding series.
However, for the higher powers, no simple result appears known
for the sum over Nielsen's functions on the unit circle.
We refer to \appref{a:generalization} for 
more details.

We now turn to some other series.
\begin{series}\label{series:s1}
\begin{equation}
\sum_{n=1}^\infty\frac{1}{n^2}\frac{[\Gamma(n)]^2}{\Gamma(2n)}
=-\frac{8}{3}\zeta(3) +\frac{4\pi}{3}\Cl{2}\left(\frac{\pi}{3}\right)
\end{equation}
\end{series}
\begin{proof}
We start by using the duplication formula for the gamma function to
get a hypergeometric series,
\begin{equation}
\begin{split}
&\sum_{n=1}^\infty\frac{1}{n^2}\frac{[\Gamma(n)]^2}{\Gamma(2n)}
=2\sum_{n=1}^\infty\frac{1}{n^2}
\frac{\Gamma\left(\frac{1}{2}\right)\Gamma(n)}
{\Gamma\left(\frac{1}{2}+n\right)}\left(\frac{1}{4}\right)^n\\
&={}_4F_3\left(1,1,1,1;\frac{3}{2},2,2;\frac{1}{4}\right)
=\int_0^1{\rm d}t\,{}_3F_2\left(1,1,1;\frac{3}{2},2;\frac{1}{4}t\right).
\end{split}
\end{equation}
In the last step we have used the integral representation for 
the function ${}_4F_3$, eq.~\eqref{e:4F3integral}.
By using the identity (7.4.2.353) of 
\cite{Prudnikov} and afterwards making  the substitution 
$\frac{\sqrt{t}}{2}=\sin \frac{u}{2}$ we get
\begin{equation*}
4\int_0^1\frac{{\rm d}t}{t}\arcsin^2\frac{\sqrt{t}}{2}
=2\int_0^{\frac{\pi}{3}}\frac{{\rm d}u}{2\tan\frac{u}{2}}u^2.
\end{equation*}
We will use integration by parts to decompose this integral. We get
\begin{equation*}
2\left.u^2\log\left(2\sin \frac{u}{2}\right)\right|_0^\frac{\pi}{3}
-4\int_0^{\frac{\pi}{3}}{\rm d}u\ u\log\left(2\sin \frac{u}{2}\right)
=-4\int_0^{\frac{\pi}{3}}{\rm d}u\ u\log\left(2\sin \frac{u}{2}\right).
\end{equation*}
Invoking eqs.~(6.52) and (6.46) of \cite{Lewin}, we get\footnote{There
is a misprint in eq.~(6.52) of \cite{Lewin}. The correct result is
\begin{equation*}
\int_0^\theta{\rm d}\theta\log\left(2\sin\frac{\theta}{2}\right)
=\zeta(3)-\theta\Cl{2}(\theta)-\Cl{3}(\theta).
\end{equation*}
}  
\begin{equation*}
-4\left\{\zeta(3)-\frac{\pi}{3}\Cl{2}\left(\frac{\pi}{3}\right)
-\Cl{3}\left(\frac{\pi}{3}\right)\right\}
=\frac{4\pi}{3}\Cl{2}\left(\frac{\pi}{3}\right)-\frac{8}{3}\zeta(3).
\end{equation*}
\end{proof}
\begin{series}\label{series:s2}
\begin{equation}
\sum_{n=1}^\infty\frac{(-1)^n}{n^2}\frac{[\Gamma(n)]^2}{\Gamma(2n)}=
-\frac{4}{5}\zeta(3)
\end{equation}
\end{series}
\begin{proof}
We may rewrite this series as
\begin{equation*}
\sum_{n=1}^\infty\frac{(-1)^n}{n^2}\frac{[\Gamma(n)]^2}{\Gamma(2n)}
=-2\sum_{n=1}^\infty\frac{(-1)^{n+1}}{n^3}\frac{[n!]^2}{(2n)!}
=-\tfrac{4}{5}\zeta(3),
\end{equation*}
where eq.~(0.1) in Chapter~9 of \cite{Berndt} gives us the sum of this series.
\end{proof}
\section{Two-dimensional series}
\label{s:two-d}
A large number of two-dimensional series encountered in 
Feynman integral calculations can also be summed analytically. 
Some results for such series will be presented here along with proofs,
many of which are based on the results of our theorem. 
The simpler ones are evaluated by summing over one variable and 
recognizing the result as a one-dimensional series for which we know 
the sum. For the other series, we will need to use integration- or 
differentiation methods to analytically evaluate the sum.

When summing  the simpler two-dimensional series, we will often find the 
following result useful,
\begin{equation}
\begin{split}
\sum_{n=1}^\infty\frac{1}{n+a}\frac{1}{n+b}
&=\frac{1}{(1+a)(1+b)}\,
{}_3F_2(1,1+a,1+b;2+a,2+b;1)\\
&=\frac{1}{b-a}\left[\psi(1+b)-\psi(1+a)\right], \quad a\neq b.
\label{e:n-sum}
\end{split}
\end{equation}
The last step follows by using eq.~(7.4.4.33) of \cite{Prudnikov}.

Another useful result is
\begin{equation}
\sum_{n=1}^\infty\frac{\Gamma(n+k)}{\Gamma(1+n+2k)}
=\frac{\Gamma(1+k)}{\Gamma(2+2k)}\,
{}_2F_1(1,1+k;2+2k;1)
=\frac{\Gamma(k)}{\Gamma(1+2k)}.
\label{e:sumgamma}
\end{equation}
This follows immediately by recognizing the series as a hypergeometric 
function and using \eqref{e:2F1unity}.

\begin{series}\label{series:s3}
\begin{equation}
\sum_{n=1}^\infty\sum_{k=1}^\infty
\frac{k}{n(1+k)^2(n+k)}=\zeta(3).
\end{equation}
\end{series}
\begin{proof}
We start by summing over $n$, using \eqref{e:n-sum}, to get
\begin{equation*}
\begin{split}
&\sum_{n=1}^\infty\sum_{k=1}^\infty
\frac{k}{n(1+k)^2(n+k)}
=\sum_{k=1}^\infty\frac{1}{(1+k)^2}[\gamma+\psi(1+k)]\\
&=\sum_{k=2}^\infty\frac{1}{k^2}[\gamma+\psi(k)]
=\sum_{k=1}^\infty\frac{1}{k^2}[\gamma+\psi(k)]=\zeta(3),
\end{split}
\end{equation*}
where we have made use of \eqref{e:psisum1} in
the last step.
\end{proof}
\begin{series}\label{series:s4}
\begin{equation}
\sum_{n=1}^\infty\sum_{k=1}^\infty
\frac{1}{k(n+k)(n+2k)}=\tfrac{3}{4}\zeta(3).
\end{equation}
\end{series}
\begin{proof}
Also here we start by summing over $n$, using \eqref{e:n-sum}, to get
\begin{equation*}
\sum_{n=1}^\infty\sum_{k=1}^\infty
\frac{1}{k(n+k)(n+2k)}
=\sum_{k=1}^\infty\frac{1}{k^2}[\psi(1+2k)-\psi(1+k)]=\tfrac{3}{4}\zeta(3),
\end{equation*}
where we have made use of \eqref{e:psisum2} and \eqref{e:psisum4} in
the last step.
\end{proof}
\begin{series}\label{series:s5}
\begin{equation}
\sum_{n=0}^\infty\sum_{k=1}^\infty
\frac{1}{k(n+k)(n+2k)}=\tfrac{5}{4}\zeta(3).
\end{equation}
\end{series}
\begin{proof}
This follows as an immediate corollary from the previous result.
We rewrite the sum as
\begin{equation*}
\begin{split}
\sum_{n=0}^\infty\sum_{k=1}^\infty
\frac{1}{k(n+k)(n+2k)}
&=\frac{1}{2}\sum_{k=1}^\infty\frac{1}{k^3}
+\sum_{n=1}^\infty\sum_{k=1}^\infty
\frac{1}{k(n+k)(n+2k)}\\
&=\half\zeta(3)+\tfrac{3}{4}\zeta(3)
=\tfrac{5}{4}\zeta(3).
\end{split}
\end{equation*}
\end{proof}
This result can also be found with a different proof in eq.~(1.2) of 
\cite{Sita} by using the identity \eqref{e:changesum} to relate these series.
\begin{series}\label{series:s6}
\begin{equation}
\sum_{n=1}^\infty\sum_{k=1}^\infty\frac{1}{k!}
\frac{\Gamma(2k)\Gamma(n+k)}{\Gamma(1+n+2k)}=\half\zeta(2).
\end{equation}
\end{series}
\begin{proof}
We start by summing over $n$, using \eqref{e:sumgamma}, to get
\begin{equation*}
\sum_{n=1}^\infty\sum_{k=1}^\infty\frac{1}{k!}
\frac{\Gamma(2k)\Gamma(n+k)}{\Gamma(1+n+2k)}
=\frac{1}{2}\sum_{k=1}^\infty\frac{1}{k^2}=\half\zeta(2).
\end{equation*}
\end{proof}
\begin{series}\label{series:s7}
\begin{equation}
\sum_{n=1}^\infty\sum_{k=1}^\infty\frac{1}{k!k}
\frac{\Gamma(2k)\Gamma(n+k)}{\Gamma(1+n+2k)}=\half\zeta(3).
\end{equation}
\end{series}
\begin{proof}
The sum over $n$ is the same as above. Thus,
\begin{equation*}
\sum_{n=1}^\infty\sum_{k=1}^\infty\frac{1}{k!k}
\frac{\Gamma(2k)\Gamma(n+k)}{\Gamma(1+n+2k)}
=\frac{1}{2}\sum_{k=1}^\infty\frac{1}{k^3}=\half\zeta(3).
\end{equation*}
\end{proof}
\begin{series}\label{series:s8}
\begin{equation}
\sum_{n=1}^\infty\sum_{k=1}^\infty\frac{1}{k!}
\frac{\Gamma(2k)\Gamma(n+k)}{\Gamma(1+n+2k)}[\gamma+\psi(k)]
=\half\zeta(3).
\end{equation}
\end{series}
\begin{proof}
Again we start by summing over $n$, using \eqref{e:sumgamma}, to get
\begin{equation*}
\sum_{n=1}^\infty\sum_{k=1}^\infty\frac{1}{k!}
\frac{\Gamma(2k)\Gamma(n+k)}{\Gamma(1+n+2k)}[\gamma+\psi(k)]
=\frac{1}{2}\sum_{k=1}^\infty\frac{1}{k^2}[\gamma+\psi(k)]
=\half\zeta(3),
\end{equation*}
where we have made use of \eqref{e:psisum1} in the last step. 
\end{proof}
\begin{series}\label{series:s9}
\begin{equation}
\sum_{n=1}^\infty\sum_{k=1}^\infty\frac{1}{k!}
\frac{\Gamma(2k)\Gamma(n+k)}{\Gamma(1+n+2k)}[\gamma+\psi(1+k)]=\zeta(3).
\end{equation}
\end{series}
\begin{proof}
This result follows as an immediate corollary of the two previous
results by using the recurrence relation
\eqref{e:psirecurrence}.
\end{proof}
\begin{series}\label{series:s10}
\begin{equation}
\sum_{n=1}^\infty\sum_{k=1}^\infty\frac{1}{k!}
\frac{\Gamma(2k)\Gamma(n+k)}{\Gamma(1+n+2k)}[\gamma+\psi(1+2k)]=
\tfrac{11}{8}\zeta(3).
\end{equation}
\end{series}
\begin{proof}
We start once more by summing over $n$, using \eqref{e:sumgamma}, to get
\begin{equation*}
\sum_{n=1}^\infty\sum_{k=1}^\infty\frac{1}{k!}
\frac{\Gamma(2k)\Gamma(n+k)}{\Gamma(1+n+2k)}[\gamma+\psi(1+2k)]
=\frac{1}{2}\sum_{k=1}^\infty\frac{1}{k^2}[\gamma+\psi(1+2k)]
=\tfrac{11}{8}\zeta(3),
\end{equation*}
where we have made use of \eqref{e:psisum4} in the last step.
\end{proof}

We now turn to similar series involving $n$ in the argument 
of the psi function.
Here, the results \eqref{e:n-sum} and \eqref{e:sumgamma} are not
sufficient. We will require the use of integration- and 
differentiation methods throughout the rest of this section.
\begin{series}\label{series:s11}
\begin{equation}
\sum_{n=1}^\infty\sum_{k=1}^\infty\frac{1}{k!}
\frac{\Gamma(2k)\Gamma(n+k)}{\Gamma(1+n+2k)}[\gamma+\psi(n)]
=\tfrac{7}{8}\zeta(3).
\end{equation}
\end{series}
\begin{proof}
The relation \eqref{e:psiintegral} is used to write
\begin{equation*}
\begin{split}
&\sum_{n=1}^\infty\sum_{k=1}^\infty\frac{1}{k!}
\frac{\Gamma(2k)\Gamma(n+k)}{\Gamma(1+n+2k)}[\gamma+\psi(n)]
=\sum_{n=1}^\infty\sum_{k=1}^\infty\frac{1}{k!}
\frac{\Gamma(2k)\Gamma(n+k)}{\Gamma(1+n+2k)}
\int_0^1{\rm d}t\frac{1-t^{n-1}}{1-t}\\
&=\frac{1}{2}\int_0^1\frac{{\rm d}t}{1-t}\sum_{k=1}^\infty\left\{
\frac{1}{k^2}-\frac{1}{k(1+2k)}{}_2F_1(1,1+k;2+2k;t)\right\},
\end{split}
\end{equation*}
where we have now interchanged the order of integration and
summation, thereby being able to perform the sum over $n$. Next, we
apply the integral representation \eqref{e:2F1integral} for ${}_2F_1$
before summing over $k$. Thus, we get
\begin{equation*}
\begin{split}
&\frac{1}{2}\int_0^1\frac{{\rm d}t}{1-t}\sum_{k=1}^\infty\left\{
\frac{1}{k^2}-\frac{1}{k}\int_0^1\frac{{\rm d}s}{1-ts}
\left[\frac{(1-s)^2}{1-ts}\right]^k\right\}\\
&=\frac{1}{2}\int_0^1\frac{{\rm d}t}{1-t}\left\{
\zeta(2)+\int_0^1\frac{{\rm d}s}{1-ts}
\log \left[1-\frac{(1-s)^2}{1-ts}\right]\right\}\\
&=\frac{1}{2}\int_0^1\frac{{\rm d}t}{1-t}\left\{
\zeta(2)+\int_0^1\frac{{\rm d}s}{1-ts}
[\log s+\log(2-t-s)-\log(1-ts)]\right\}\\
&=\frac{1}{2}\int_0^1\frac{{\rm d}t}{1-t}\left\{
\zeta(2)+\int_0^1\frac{{\rm d}s\log s}{1-ts}
+\int_0^1\frac{{\rm d}s\log(2-t-s)}{1-ts}
-\int_0^1\frac{{\rm d}s\log(1-ts)}{1-ts}\right\}.
\end{split}
\end{equation*}
Using eqs.~(3.12.1) and (3.14.5) of \cite{D&D} along with
eq.~(A.3.1.3) of \cite{Lewin} to perform the integration over $s$, we get
\begin{equation*}
\begin{split}
&\frac{1}{2}\int_0^1\frac{{\rm d}t}{1-t}\left\{
\zeta(2)-\frac{1}{t}\left[\Li{2}(t)
+\frac{1}{2}\log^2\left(\frac{1-t}{t}\right)
-\frac{1}{2}\log^2\left(\frac{1}{t}\right)\right.\right.\\
&\phantom{\frac{1}{2}\int_0^1\frac{{\rm d}t}{1-t}\biggl\{}
\left.\left.+\Li{2}(1-t)-\Li{2}[(1-t)^2]-\frac{1}{2}\log^2(1-t)\right]\right\}.
\end{split}
\end{equation*}
The identity (2.2.1) of \cite{D&D} is used to rewrite this as
\begin{equation*}
\begin{split}
&\frac{1}{2}\int_0^1\frac{{\rm d}t}{1-t}\left\{
\zeta(2)+\frac{1}{t}\left[\Li{2}[(1-t)^2]+2\log(t)\log(1-t)
-\zeta(2)\right]\right\}\\
&=\frac{1}{2}\int_0^1\frac{{\rm d}t}{t}\left\{
\zeta(2)+\frac{1}{1-t}\left[\Li{2}(t^2)+2\log(t)\log(1-t)
-\zeta(2)\right]\right\}\\
&=\int_0^1\frac{{\rm d}t}{t(1-t)}\log(t)\log(1-t)
+\frac{1}{2}\int_0^1\frac{{\rm d}t}{t(1-t)}
\left[\Li{2}(t^2)-t\, \zeta(2)\right],
\end{split}
\end{equation*}
where we now have substituted $t\to 1-t$ in the second step. By
expanding in partial fractions, we arrive at
\begin{equation*}
\begin{split}
&\int_0^1\frac{{\rm d}t}{t}\log(t)\log(1-t)
+\int_0^1\frac{{\rm d}t}{1-t}\log(t)\log(1-t)\\
&+\frac{1}{2}\int_0^1\frac{{\rm d}t}{t}
\left[\Li{2}(t^2)-t\zeta(2)\right]
+\frac{1}{2}\int_0^1\frac{{\rm d}t}{1-t}
\left[\Li{2}(t^2)-t\, \zeta(2)\right]\\
&=2\int_0^1\frac{{\rm d}t}{t}\log(t)\log(1-t)
+\int_0^1\frac{{\rm d}t}{t}
\left[\Li{2}(t)+\Li{2}(-t)\right]-\half\zeta(2)\\
&+\int_0^1\frac{{\rm d}t}{1-t}
\left[\Li{2}(t)+\Li{2}(-t)-\half\zeta(2)\right]
+\half\zeta(2).
\end{split}
\end{equation*}
The first two integrals have been combined by substituting $t\to 1-t$ 
in the second one. Identity (2.2.8) of \cite{D&D} has been used
to transform the two last integrals. 
In order to proceed, the first two of these integrals are
evaluated using eqs. (3.6.21), (3.8.13) and (3.8.15) of
\cite{D&D}. The result is
\begin{equation*}
\tfrac{9}{4}\zeta(3)+\int_0^1\frac{{\rm d}t}{1-t}
\left[\Li{2}(t)-\zeta(2)\right]
+\int_0^1\frac{{\rm d}t}{1-t}
\left[\Li{2}(-t)+\half\zeta(2)\right]=\tfrac{7}{8}\zeta(3),
\end{equation*}
where eqs.~(3.8.9) and (3.8.11) of \cite{D&D} have been
used\footnote{Note: There is a sign error in eq. (3.8.11) of
  \cite{D&D}. The correct result is
\begin{equation*}
\int_0^1\frac{{\rm d}t}{1-t}\left[\Li{2}(-t)+\half\zeta(2)\right]
=\tfrac{5}{8}\zeta(3).
\end{equation*}
} 
in the last step.
\end{proof}
\begin{series}\label{series:s12}
\begin{equation}
\sum_{n=1}^\infty\sum_{k=1}^\infty\frac{1}{k!}
\frac{\Gamma(2k)\Gamma(n+k)}{\Gamma(1+n+2k)}[\gamma+\psi(n+1)]
=\tfrac{5}{4}\zeta(3).
\end{equation}
\end{series}
\begin{proof}
The recurrence relation \eqref{e:psirecurrence} for the psi function
is used to write 
\begin{equation*}
\begin{split}
&\sum_{n=1}^\infty\sum_{k=1}^\infty\frac{1}{k!}
\frac{\Gamma(2k)\Gamma(n+k)}{\Gamma(1+n+2k)}[\gamma+\psi(n+1)]\\
&=\sum_{n=1}^\infty\sum_{k=1}^\infty\frac{1}{k!}
\frac{\Gamma(2k)\Gamma(n+k)}{\Gamma(1+n+2k)}[\gamma+\psi(n)]
+\sum_{n=1}^\infty\sum_{k=1}^\infty\frac{1}{k!}
\frac{\Gamma(2k)\Gamma(n+k)}{\Gamma(1+n+2k)}\frac{1}{n}\\
&=\tfrac{7}{8}\zeta(3)+\frac{1}{2}\sum_{k=1}^\infty\frac{1}{k(1+2k)}
{}_3F_2(1,1,1+k;2,2+2k;1),
\end{split}
\end{equation*}
where now the sum over $n$ has been performed in the last term. By
using eq. (7.4.4.40) of \cite{Prudnikov}, we get
\begin{equation*}
\tfrac{7}{8}\zeta(3)+\frac{1}{2}\sum_{k=1}^\infty\frac{1}{k^2} 
\left\{\psi(1+2k)-\psi(1+k)\right\}=\tfrac{5}{4}\zeta(3),
\end{equation*}
where in the last step, eqs.~\eqref{e:psisum2} and 
\eqref{e:psisum4} have been used.
\end{proof}
\begin{series}\label{series:s13}
\begin{equation}
\sum_{n=1}^\infty\sum_{k=1}^\infty\frac{1}{k!}
\frac{\Gamma(2k)\Gamma(n+k)}{\Gamma(1+n+2k)}[\gamma+\psi(n+k)]
=\tfrac{3}{2}\zeta(3).
\end{equation}
\end{series}
\begin{proof}
We can rewrite the sum as
\begin{equation*}
\begin{split}
&\sum_{n=1}^\infty\sum_{k=1}^\infty\frac{1}{k!}
\frac{\Gamma(2k)\Gamma(n+k)}{\Gamma(1+n+2k)}[\gamma+\psi(n+k)]\\
&=\gamma\sum_{n=1}^\infty\sum_{k=1}^\infty\frac{1}{k!}
\frac{\Gamma(2k)\Gamma(n+k)}{\Gamma(1+n+2k)}
+\left.\frac{{\rm d}}{{\rm d}x}\right|_{x=0}
\sum_{n=1}^\infty\sum_{k=1}^\infty\frac{1}{k!}
\frac{\Gamma(2k)\Gamma(n+k+x)}{\Gamma(1+n+2k)}\\
&=\frac{\gamma}{2}\sum_{k=1}^\infty\frac{1}{k^2}
+\left.\frac{{\rm d}}{{\rm d}x}\right|_{x=0}
\sum_{k=1}^\infty\frac{\Gamma(1+k+x)\Gamma(2k)}{k!\Gamma(2+2k)}
{}_2F_1(1,1+k+x;2+2k;1)\\
&=\frac{\gamma}{2}\sum_{k=1}^\infty\frac{1}{k^2}
+\left.\frac{{\rm d}}{{\rm d}x}\right|_{x=0}
\sum_{k=1}^\infty\frac{\Gamma(1+k+x)\Gamma(2k)}{k!(k-x)\Gamma(1+2k)}\\
&=\frac{1}{2}\sum_{k=1}^\infty\frac{1}{k^2}
\left[\frac{1}{k}+\gamma+\psi(1+k)\right]\\
&=\half\zeta(3)
+\frac{1}{2}\sum_{k=1}^\infty\frac{1}{k^2}[\gamma+\psi(1+k)]
=\tfrac{3}{2}\zeta(3),
\end{split}
\end{equation*}
where the result \eqref{e:psisum2} has been used in the last step.
\end{proof}
\begin{series}\label{series:s14}
\begin{equation}
\sum_{n=1}^\infty\sum_{k=1}^\infty\frac{1}{k!}
\frac{\Gamma(2k)\Gamma(n+k)}{\Gamma(1+n+2k)}[\gamma+\psi(1+n+2k)]
=\tfrac{15}{8}\zeta(3).
\end{equation}
\end{series}
\begin{proof}
We can rewrite this series as
\begin{equation*}
\begin{split}
&\sum_{n=1}^\infty\sum_{k=1}^\infty\frac{1}{k!}
\frac{\Gamma(2k)\Gamma(n+k)}{\Gamma(1+n+2k)}[\gamma+\psi(1+n+2k)]\\
&=\gamma\sum_{n=1}^\infty\sum_{k=1}^\infty\frac{1}{k!}
\frac{\Gamma(2k)\Gamma(n+k)}{\Gamma(1+n+2k)}
+\left.\frac{{\rm d}}{{\rm d}x}\right|_{x=0}
\sum_{n=1}^\infty\sum_{k=1}^\infty\frac{1}{k!}
\frac{\Gamma(2k)\Gamma(n+k)}{\Gamma(1+n+2k-x)}\\
&=\frac{\gamma}{2}\sum_{k=1}^\infty\frac{1}{k^2}
+\left.\frac{{\rm d}}{{\rm d}x}\right|_{x=0}
\sum_{k=1}^\infty\frac{\Gamma(2k)}{\Gamma(2+2k-x)}
{}_2F_1(1,1+k;2+2k-x;1)\\
&=\frac{\gamma}{2}\sum_{k=1}^\infty\frac{1}{k^2}
+\left.\frac{{\rm d}}{{\rm d}x}\right|_{x=0}
\sum_{k=1}^\infty\frac{\Gamma(2k)}{(k-x)\Gamma(1+2k-x)}\\
&=\frac{1}{2}\sum_{k=1}^\infty\frac{1}{k^2}
\left[\frac{1}{k}+\gamma+\psi(1+2k)\right]\\
&=\frac{1}{2}\zeta(3)
+\frac{1}{2}\sum_{k=1}^\infty\frac{1}{k^2}[\gamma+\psi(1+2k)]
=\tfrac{15}{8}\zeta(3),
\end{split}
\end{equation*}
where the result \eqref{e:psisum4} has been used in the last step.
\end{proof}
\begin{series}\label{series:s15}
\begin{equation}
\sum_{n=0}^\infty\sum_{k=1}^\infty\frac{(-1)^k}{n!k!}\frac{1}{n+k}
\frac{[\Gamma(n+k)]^3}{\Gamma(2n+2k)}
=\frac{8}{3}\zeta(3)-\frac{4\pi}{3}\Cl{2}\left(\frac{\pi}{3}\right).
\end{equation}
\end{series}
\begin{proof}
First, we use the duplication formula for the gamma function and perform 
the sum over $n$ to get a hypergeometric function,
\begin{equation*}
\begin{split}
&\sum_{n=0}^\infty\sum_{k=1}^\infty\frac{(-1)^k}{n!k!}\frac{1}{n+k}
\frac{[\Gamma(n+k)]^3}{\Gamma(2n+2k)}\\
&=2\sum_{n=0}^\infty\sum_{k=1}^\infty\frac{(-1)^k}{n!k!}\frac{1}{n+k}
\frac{[\Gamma(n+k)]^2\Gamma\left(\frac{1}{2}\right)}
{\Gamma\left(\frac{1}{2}+n+k\right)}\left(\frac{1}{4}\right)^{n+k}\\
&=2\sum_{k=1}^\infty\frac{\Gamma(k)\Gamma\left(\frac{1}{2}\right)}
{k^2\Gamma\left(\frac{1}{2}+k\right)}
{}_3F_2\left(k,k,k;k+1,k+\frac{1}{2};\frac{1}{4}\right)
\left(-\frac{1}{4}\right)^k.
\end{split}
\end{equation*}
By using the integral representation \eqref{e:3F2integral2},
we are able to perform the sum over $k$,
\begin{equation*}
\begin{split}
&2\sum_{k=1}^\infty\frac{1}{k}\int_0^1{\rm d}s\int_0^1{\rm d}t\ s^{k-1}t^{k-1}
(1-t)^{-\frac{1}{2}}\left(1-\frac{1}{4}st\right)^{-k}
\left(-\frac{1}{4}\right)^k\\
&=2\int_0^1{\rm d}s\int_0^1{\rm d}t
\frac{1}{st\sqrt{1-t}}\sum_{k=1}^\infty\frac{1}{k}
\left(\frac{-st}{4-st}\right)^k\\
&=-2\int_0^1{\rm d}s\int_0^1{\rm d}t\frac{1}{st\sqrt{1-t}}
\log\left(1+\frac{st}{4-st}\right)\\
&=2\int_0^1{\rm d}s\int_0^1{\rm d}t\frac{1}{st\sqrt{1-t}}
\log\left(1-\frac{st}{4}\right).
\end{split}
\end{equation*}
Next, we integrate over $s$, and thereafter
introduce the series representation of the dilogarithm:
\begin{equation*}
\begin{split}
&-2\int_0^1{\rm d}t\frac{1}{t\sqrt{1-t}}
\Li{2}\left(\frac{t}{4}\right)
=-2\int_0^1{\rm d}t\frac{1}{t\sqrt{1-t}}
\sum_{n=1}^\infty\frac{t^n}{n^2}\left(\frac{1}{4}\right)^n\\
&=-2\sum_{n=1}^\infty\frac{1}{n^2}\left(\frac{1}{4}\right)^n
\int_0^1{\rm d}t\ t^{-1+n}(1-t)^{-\frac{1}{2}}\\
&=-2\sum_{n=1}^\infty\frac{1}{n^2}
\frac{\Gamma(n)\Gamma\left(\frac{1}{2}\right)}
{\Gamma\left(\frac{1}{2}+n\right)}\left(\frac{1}{4}\right)^n
=\frac{8}{3}\zeta(3)-\frac{4\pi}{3}\Cl{2}\left(\frac{\pi}{3}\right).
\end{split}
\end{equation*}
In the last two steps, we performed the integration over $t$
and arrived at a series encountered in the proof of \seriesref{series:s1}.
\end{proof}
\begin{series}\label{series:s16}
\begin{equation}
\sum_{n=0}^\infty\sum_{k=1}^\infty\frac{1}{n!k!}\frac{1}{n+k}
\frac{[\Gamma(n+k)]^3}{\Gamma(2n+2k)}
=-\frac{41}{24}\zeta(3)-\frac{4\pi}{3}\Cl{2}\left(\frac{\pi}{3}\right)
+\frac{\pi^2}{4}\log 2 +2\pi G.
\end{equation}
\end{series}
\begin{proof}
The sum over $n$ is the same as for the previous case,
\begin{equation*}
\begin{split}
&\sum_{n=0}^\infty\sum_{k=1}^\infty\frac{1}{n!k!}\frac{1}{n+k}
\frac{[\Gamma(n+k)]^3}{\Gamma(2n+2k)}\\
&=2\sum_{k=1}^\infty\frac{\Gamma(k)\Gamma\left(\frac{1}{2}\right)}
{k^2\Gamma\left(k+\frac{1}{2}\right)}
{}_3F_2\left(k,k,k;k+1,k+\frac{1}{2};\frac{1}{4}\right)
\left(\frac{1}{4}\right)^k.
\end{split}
\end{equation*}
By using the integral representation \eqref{e:3F2integral2}
of ${}_3F_2$, we are able to 
perform the sum over $k$,
\begin{equation*}
\begin{split}
&2\sum_{k=1}^\infty\frac{1}{k}\int_0^1{\rm d}s
\int_0^1{\rm d}t\ s^{k-1}t^{k-1}
(1-t)^{-\frac{1}{2}}\left(1-\frac{1}{4}st\right)^{-k}
\left(\frac{1}{4}\right)^k\\
&=2\int_0^1{\rm d}s\int_0^1{\rm d}t
\frac{1}{st\sqrt{1-t}}\sum_{k=1}^\infty\frac{1}{k}
\left(\frac{st}{4-st}\right)^k\\
&=-2\int_0^1{\rm d}s\int_0^1{\rm d}t\frac{1}{st\sqrt{1-t}}
\log\left(1-\frac{st}{4-st}\right)\\
&=2\int_0^1{\rm d}s\int_0^1{\rm d}t\frac{1}{st\sqrt{1-t}}
\left\{\log\left(1-\frac{st}{4}\right)
-\log\left(1-\frac{st}{2}\right)\right\}\\
&=-2\int_0^1{\rm d}t\frac{1}{t\sqrt{1-t}}\, 
\Li{2}\left(\frac{t}{4}\right)
+2\int_0^1{\rm d}t\frac{1}{t\sqrt{1-t}}\,
\Li{2}\left(\frac{t}{2}\right).
\end{split}
\end{equation*}
The first of these two integrals is known from the previous proof, 
while the second one is calculated in the same way as we did 
for the first one. Thus, we get
\begin{equation*}
\begin{split}
&\frac{8}{3}\zeta(3)-\frac{4\pi}{3}\Cl{2}\left(\frac{\pi}{3}\right)
+2\int_0^1{\rm d}t\frac{1}{t\sqrt{1-t}}\sum_{n=1}^\infty\frac{t^n}{n^2}
\left(\frac{1}{2}\right)^n\\
&=\frac{8}{3}\zeta(3)-\frac{4\pi}{3}\Cl{2}\left(\frac{\pi}{3}\right)
+2\sum_{n=1}^\infty\frac{1}{n^2}\left(\frac{1}{2}\right)^n
\int_0^1{\rm d}t\ t^{-1+n}(1-t)^{-\frac{1}{2}}\\
&=\frac{8}{3}\zeta(3)-\frac{4\pi}{3}\Cl{2}\left(\frac{\pi}{3}\right)
+2\sum_{n=1}^\infty\frac{1}{n^2}
\frac{\Gamma(n)\Gamma(\half)}{\Gamma(\half+n)}\\
&=\frac{8}{3}\zeta(3)-\frac{4\pi}{3}\Cl{2}\left(\frac{\pi}{3}\right)
+2\ {}_4F_3\left(1,1,1,1;\frac{3}{2},2,2;\frac{1}{2}\right)\\
&=\frac{8}{3}\zeta(3)-\frac{4\pi}{3}\Cl{2}\left(\frac{\pi}{3}\right)
+2\int_0^1{\rm d}t\ {}_3F_2\left(1,1,1;\frac{3}{2},2;
\frac{1}{2}t\right),
\end{split}
\end{equation*}
where we have used the integral representation for ${}_4F_3$. 
Next, we use 
eq. (7.4.2.353) of \cite{Prudnikov} to get
\begin{equation*}
\frac{8}{3}\zeta(3)-\frac{4\pi}{3}\Cl{2}\left(\frac{\pi}{3}\right)
+4\int_0^1\frac{{\rm d}t}{t}\arcsin^2\sqrt{\frac{t}{2}}.
\end{equation*}
The substitution $\sqrt{\frac{t}{2}}=\sin\frac{w}{2}$ yields
\begin{equation*}
\frac{8}{3}\zeta(3)-\frac{4\pi}{3}\Cl{2}\left(\frac{\pi}{3}\right)
+2\int_0^\frac{\pi}{2}\frac{{\rm d}w}{2\tan\frac{w}{2}}w^2.
\end{equation*}
To evaluate this integral, we integrate by parts to get
\begin{equation*}
\begin{split}
&\frac{8}{3}\zeta(3)-\frac{4\pi}{3}\Cl{2}\left(\frac{\pi}{3}\right)
+\left.2w^2\log\left[2\sin\frac{w}{2}\right]\right|_0^{\frac{\pi}{2}}
-4\int_0^\frac{\pi}{2}{\rm d}w\ w\log\left[2\sin\frac{w}{2}\right]\\
&=\frac{8}{3}\zeta(3)-\frac{4\pi}{3}\Cl{2}\left(\frac{\pi}{3}\right)
+\frac{\pi^2}{4}\log 2 
-4\int_0^\frac{\pi}{2}{\rm d}w\ w\log\left[2\sin\frac{w}{2}\right]\\
&=\frac{8}{3}\zeta(3)-\frac{4\pi}{3}\Cl{2}\left(\frac{\pi}{3}\right)
+\frac{\pi^2}{4}\log 2-\frac{35}{8}\zeta(3)+2\pi G\\
&=-\frac{41}{24}\zeta(3)-\frac{4\pi}{3}\Cl{2}\left(\frac{\pi}{3}\right)
+\frac{\pi^2}{4}\log 2 +2\pi G,
\end{split}
\end{equation*}
by using eq.~(6.52) of \cite{Lewin} in the last step. 
\end{proof}
\begin{series}\label{series:s17}
\begin{equation}
\sum_{n=1}^\infty\sum_{k=1}^\infty\frac{1}{k!}\frac{1}{n(n+k)}
\frac{\Gamma(n+k)\Gamma(n+2k)}{\Gamma(2n+2k)}
=\frac{47}{12}\zeta(3)-\frac{4\pi}{3}\Cl{2}\left(\frac{\pi}{3}\right)
\end{equation}
\end{series}
To be able to prove this result, we will need the following lemma:
\begin{lemma}
\begin{equation}
\sum_{n=1}^k\frac{1}{n}\frac{\Gamma(2k-n)}{\Gamma(1+k-n)}=
\frac{\Gamma(2k)}{\Gamma(1+k)}\left[\psi(2k)-\psi(k)\right]
\end{equation}
\end{lemma}

\begin{proof}[Proof of the lemma]
First, we rewrite the series as
\begin{equation}
\lim_{y\to 0}\frac{1}{\Gamma(1-k+y)}\sum_{n=1}^k\frac{1}{n}
\frac{\Gamma(2k-n-y)\Gamma(1-k+y)}{\Gamma(1+k-n)}.
\label{e:lemma1}
\end{equation}
The combination of gamma functions inside the sum is a beta function 
which we will represent as an integral,
\begin{equation*}
\frac{\Gamma(2k-n-y)\Gamma(1-k+y)}{\Gamma(1+k-n)}
=\int_0^1{\rm d}t\ t^{-1+2k-n-y}(1-t)^{-k+y},
\end{equation*}
where now $k-1<\Re\ y<2k-n$ for the integral to converge. 
The factor $1/n$ is also represented as an integral,
\begin{equation*} 
\frac{1}{n}=\int_0^1{\rm d}x\ x^{n-1}.
\end{equation*}
The sum in \eqref{e:lemma1} can then be rewritten as
\begin{equation*}
\begin{split}
&\sum_{n=1}^k\frac{1}{n}
\frac{\Gamma(2k-n-y)\Gamma(1-k+y)}{\Gamma(1+k-n)}\\
&=\sum_{n=1}^k\int_0^1{\rm d}x\ x^{n-1}
\int_0^1{\rm d}t\ t^{-1+2k-n-y}(1-t)^{-k+y}\\
&=\int_0^1{\rm d}x\int_0^1{\rm d}t\ t^{-2+2k-y}(1-t)^{-k+y}
\sum_{n=1}^k\left(\frac{x}{t}\right)^{n-1}\\
&=\int_0^1{\rm d}x\int_0^1{\rm d}t\ t^{-2+2k-y}(1-t)^{-k+y}
\left[\frac{1-\left(x/t\right)^k}{1-\left(x/t\right)}\right]\\
&=\int_0^1{\rm d}x\ x^{k-1}\int_0^1{\rm d}t\ t^{-1+k-y}(1-t)^{-k+y}
\left[\frac{1-\left(t/x\right)^k}{1-\left(t/x\right)}\right].
\end{split}
\end{equation*}
For the purpose of integrating over $t$,
we would like to split the integral into two parts,
according to the two terms in the numerator.
In order to do so, avoiding the singularity from the denominator,
we introduce a shift $i\epsilon$ in the denominator.
Thus,
\begin{equation*}
\begin{split}
&\int_0^1{\rm d}x\ x^{k-1}\int_0^1{\rm d}t\ t^{-1+k-y}(1-t)^{-k+y}
\left[\frac{1-\left(t/x\right)^k}{1-\left(t/x\right)+i\epsilon}\right]\\
&=\int_0^1{\rm d}x\ x^{k-1}\left\{\Gamma(k-y)\Gamma(1-k+y)
{}_2F_1\left(1,k-y;1;\frac{1}{x}-i\epsilon\right)\right.\\
&\phantom{=\int_0^1{\rm d}x\ x^{k-1}\biggl\{}
\left.-x^{-k}\frac{\Gamma(2k-y)\Gamma(1-k+y)}{\Gamma(1+k)}
{}_2F_1\left(1,2k-y;1+k;\frac{1}{x}-i\epsilon\right)\right\}.
\end{split}
\end{equation*}
The two hypergeometric functions may be combined using
the transformation formula (7.3.1.6) of \cite{Prudnikov} 
with $a=1$, $b=1-k$, 
$c=2-2k+y$ and $z=x+i\epsilon$. Thus, we may let $\epsilon\to 0$, to get
\begin{equation*}
\begin{split}
&\frac{\Gamma(1-k+y)\Gamma(2k-1-y)}{\Gamma(k)}\int_0^1{\rm d}x\ 
{}_2F_1(1,1-k;2-2k+y;x)\\
&=\frac{\Gamma(1-k+y)\Gamma(2k-1-y)}{\Gamma(k)}\,
{}_3F_2(1,1,1-k;2,2-2k+y;1)\\
&=\frac{\Gamma(1-k+y)\Gamma(2k-y)}{\Gamma(1+k)}
\left[\psi(1-2k+y)-\psi(1-k+y)\right]\\
&=\frac{\Gamma(1-k+y)\Gamma(2k-y)}{\Gamma(1+k)}
\bigl\{\psi(2k-y)-\psi(k-y)\\
&\phantom{=\frac{\Gamma(1-k+y)\Gamma(2k-y)}{\Gamma(1+k)}\bigl\{}
+\pi\cot[\pi(2k-y)]-\pi\cot[\pi(k-y)]\bigr\}\\
&=\frac{\Gamma(1-k+y)\Gamma(2k-y)}{\Gamma(1+k)}
\left[\psi(2k-y)-\psi(k-y)\right],
\end{split}
\end{equation*}
where in the first two steps we have used eqs.~(7.2.3.9) and (7.4.4.40)
of \cite{Prudnikov}.
We have also used the reflection formula,
eq.~(6.3.7) of \cite{Abramowitz}, for the psi function.
Thus, we have shown that 
\begin{equation*}
\sum_{n=1}^k\frac{1}{n}
\frac{\Gamma(2k-n-y)}{\Gamma(1+k-n)}=
\frac{\Gamma(2k-y)}{\Gamma(1+k)}
\left[\psi(2k-y)-\psi(k-y)\right],
\end{equation*}
valid for $k-1<\Re\ y<k$. By analytic continuation, the result can be
extended to all values of $y$, except $y=k, k+1, k+2,\dotsc$.
By letting $y\to 0$, we find that
\begin{equation*}
\sum_{n=1}^k\frac{1}{n}\frac{\Gamma(2k-n)}{\Gamma(1+k-n)}=
\frac{\Gamma(2k)}{\Gamma(1+k)}\left[\psi(2k)-\psi(k)\right],
\end{equation*}
and the lemma has been proven.
\end{proof}

\begin{proof}[Proof of the series]
First, we rewrite the series as
\begin{equation*}
\begin{split}
&\sum_{n=1}^\infty\sum_{k=1}^\infty\frac{1}{k!}\frac{1}{n(n+k)}
\frac{\Gamma(n+k)\Gamma(n+2k)}{\Gamma(2n+2k)}\\
&=\sum_{n=1}^\infty\sum_{k=0}^\infty\frac{1}{k!}\frac{1}{n(n+k)}
\frac{\Gamma(n+k)\Gamma(n+2k)}{\Gamma(2n+2k)}
-\sum_{k=1}^\infty\frac{1}{k^2}
\frac{[\Gamma(k)]^2}{\Gamma(2k)}.
\end{split}
\end{equation*}
The second of these two series is already known 
(cf.~\seriesref{series:s1}), 
whereas for the two-dimensional one, we will rewrite it using the identity
\begin{equation}
\sum_{k=0}^\infty\sum_{n=1}^\infty a_{n,k}=
\sum_{k=1}^\infty\sum_{n=1}^ka_{n,k-n},\label{e:changesum}
\end{equation}
which is easily derived from identities given in chapter 4.1 of 
\cite{Hansen}. We get
\begin{equation*}
\sum_{k=1}^\infty\sum_{n=1}^k\frac{1}{nk}
\frac{\Gamma(k)\Gamma(2k-n)}{\Gamma(2k)\Gamma(1+k-n)}
-\frac{4\pi}{3}\Cl{2}\left(\frac{\pi}{3}\right)+\frac{8}{3}\zeta(3).
\end{equation*}
By the lemma, this equals 
\begin{equation*}
\sum_{k=1}^\infty\frac{1}{k^2}\left[\psi(2k)-\psi(k)\right]
-\frac{4\pi}{3}\Cl{2}\left(\frac{\pi}{3}\right)+\frac{8}{3}\zeta(3)
=\frac{47}{12}\zeta(3)-\frac{4\pi}{3}\Cl{2}\left(\frac{\pi}{3}\right),
\end{equation*}
where we have used \eqref{e:psisum1} and \eqref{e:psisum3}.
Thus, the proof is complete.
\end{proof}

It is interesting to note that the sums that appear in moderate-order
Feynman integral calculations apparently can all be expressed 
in terms of known constants.
This suggests that such integrals in some sense are of limited
`complexity'. For a qualification of this statement, we refer to
the appendix of \cite{Borwein-preprint}.
\appendix
\section{Some special functions and identities}
\label{a:specfunc}
Below, we collect some definitions and properties of special functions
that are frequently used in the proofs.

\subsection*{The psi function}
The psi function is defined as the logarithmic derivative of the
gamma function, 
$\psi(z)=({\rm d}/{\rm d}z)\, \log\Gamma(z)=\Gamma'(z)/\Gamma(z)$,
and has the following integral representation (cf.\ eq.~(6.3.22) of
\cite{Abramowitz}),
\begin{equation}
\gamma+\psi(z)=\int_0^1{\rm d}t\, \frac{1-t^{z-1}}{1-t},
\label{e:psiintegral}
\end{equation}
where $\gamma=0.577\; 216\dotsc$ is Euler's constant, and $\psi(1)=-\gamma$.
From the integral representation, one immediately finds that in the case
of positive integer arguments, we have
\begin{equation}
\gamma+\psi(n)=\sum_{j=1}^{n-1}\frac{1}{j}.
\end{equation}
The psi function satisfies the following recurrence relation (see
eq. (6.3.5) of \cite{Abramowitz}),
\begin{equation}
\psi(1+z)=\psi(z)+\frac{1}{z}.\label{e:psirecurrence}
\end{equation}

\subsection*{Polylogarithms}
Each of the polylogarithm functions can be represented as a series,
\begin{equation}
\Li{n}(z)=\sum_{k=1}^\infty\frac{z^k}{k^n}, \quad
\abs{z}<1,\quad n=0,1,2,\dotsc,
\label{e:polylogsum}
\end{equation}
or as an integral,
\begin{align}
\Li{n}(z)&=\frac{(-1)^{n-1}}{(n-2)!}\int_0^1{\rm d}t\, 
\frac{\log^{n-2}(t)\log(1-tz)}{t}\label{e:polylogintegral}
,\quad n=2,3,4,\dotsc\\
&=\int_0^z{\rm d}t\, \frac{\Li{n-1}(t)}{t},\quad n=1,2,3,\dotsc,
\label{e:polylogintegral2}
\end{align}
with $\Li{n}(1)=\zeta(n),\ n=2,3,4,\dotsc$, and
$\Li{n}(-1)=(1/2^{n-1}-1)\zeta(n),\ n=2,3,4,\dotsc$.
The following factorization formula (eq.~(7.41) of \cite{Lewin})
is valid for the polylogarithms,
\begin{equation}
\Li{n}(t^k)=k^{n-1}\sum_{j=1}^k\Li{n}(\omega^j t),\quad n=1,2,3,\dotsc,
\label{e:polylogfactorization}
\end{equation}
where $\omega=e^{2\pi i/k}$.
These and further properties of the polylogarithms and related functions 
can be found in \cite{Lewin,D&D}. 

\subsection*{Nielsen's generalized polylogarithms}
The functions $S_{n,p}(z)$ are Nielsen's generalized polylogarithms. We will 
make use of these functions for $p=2$, in which case they are given by the 
following integral representations \cite{D&D},
\begin{align}
S_{n,2}(z)&=\frac{(-1)^{n-1}}{2(n-1)!}\int_0^1{\rm d}t\,
\frac{\log^{n-1}(t)\log^2(1-zt)}{t},\quad n=1,2,3,\dotsc
\label{e:S12integral}\\
&=\int_0^z{\rm d}t\,\frac{S_{n-1,2}(t)}{t},\quad n=1,2,3,\dotsc,
\label{e:S12integral2}
\end{align}
where it is understood that $S_{0,2}(z)=\tfrac{1}{2}\log^2(1-z)$.
We shall be needing the following identity for $S_{1,2}(z)$,
\begin{equation}
S_{1,2}(z)+S_{1,2}\left(\frac{1}{z}\right)
=\Li{3}(z)
-\frac{1}{6}\log^3(-z)-\log(-z)\Li{2}(z)+\zeta(3),
\label{e:S12identity}
\end{equation}
given in eq.~(2.2.15) of \cite{D&D}.
We will use the convention $\log(-1)=i\pi$. Thus, the formula
is seen to be valid also for $z=1$.
Further identities regarding the functions $S_{n,p}(z)$ can be found in 
\cite{D&D}. Nielsen's original work is found in \cite{Nielsen}.

\subsection*{Clausen's functions}
Each of Clausen's functions can be represented as a series 
(see \cite{Lewin}),
\begin{equation}
\begin{split}
\Cl{2n}(\theta)=\sum_{k=1}^\infty\frac{\sin k\theta}{k^{2n}}, \\
\Cl{2n-1}(\theta)=\sum_{k=1}^\infty\frac{\cos k\theta}{k^{2n-1}},
\label{e:clausensum}
\end{split}
\end{equation}
where $n$ can be any positive integer.
Clausen's functions are seen to be periodic with 
period $2\pi$. 

We shall in this article need only the functions $\Cl{1}(\theta)$, 
$\Cl{2}(\theta)$ and $\Cl{3}(\theta)$.
From the definition, we see that the function $\Cl{2}(\theta)$ is 
antisymmetric, $\Cl{2}(-\theta)=-\Cl{2}(\theta)$. When the argument is an 
integer multiple of $\pi$, the function vanishes, $\Cl{2}(k\pi)=0$.

On the unit circle, the imaginary part of the dilogarithm is 
Clausen's function $\Cl{2}(\theta)$,
\begin{equation}
\Im \left\{\Li{2}(e^{i\theta})\right\}=\Cl{2}(\theta).
\end{equation}
This function satisfies the factorization formula
\begin{equation}
\frac{1}{2}\Cl{2}(2\theta)=\Cl{2}(\theta)-\Cl{2}(\pi-\theta).
\label{e:clausenfactorization}
\end{equation}
The Clausen function $\Cl{2}(\theta)$ has its maximum value for $\theta=\pi/3$,
$\Cl{2}(\pi/3)=1.014\; 942\dotsc$. Furthermore,
\begin{equation}
\Cl{2}\left(\frac{2\pi}{3}\right)
=\frac{2}{3}\Cl{2}\left(\frac{\pi}{3}\right), \qquad
\Cl{2}\left(\frac{\pi}{2}\right)
=G,\label{e:clausenhalfpi}
\end{equation}
where $G=0.915\; 966\dotsc$ is the Catalan constant.

\subsection*{Hypergeometric functions}
We will frequently use the series definitions of the following 
hypergeometric functions,
\begin{gather}
{}_2F_1(a,b;c;z)=\sum_{n=0}^\infty\frac{(a)_n(b)_n}{(c)_n}\frac{z^n}{n!},\\
{}_3F_2(a,b,c;d,e;z)=\sum_{n=0}^\infty\frac{(a)_n(b)_n(c)_n}{(d)_n(e)_n}
\frac{z^n}{n!},\\
{}_4F_3(a,b,c,d;e,f,g;z)=\sum_{n=0}^\infty
\frac{(a)_n(b)_n(c)_n(d)_n}{(e)_n(f)_n(g)_n}
\frac{z^n}{n!},
\end{gather}
where $(a)_n$ is the Pochhammer symbol,
\begin{equation*}
(a)_n\equiv\frac{\Gamma(a+n)}{\Gamma(a)}
=a(a+1)\dotsm(a+n-1).
\end{equation*}
It is evident that these functions are all symmetric with
respect to the ``numerator'' or the ``denominator'' arguments.
All these series converge for $\abs{z}<1$. Conditions for convergence on the
unit circle are given in Chapter~7 of \cite{Prudnikov}.
 
These functions all have various integral representations. We shall be 
using the following ones,
\begin{align}
{}_2F_1(a,b;c;z)&=\frac{\Gamma(c)}{\Gamma(b)\Gamma(c-b)}
\int_0^1{\rm d}t\,t^{b-1}(1-t)^{c-b-1}(1-tz)^{-a}\label{e:2F1integral},\\
{}_3F_2(a,b,c;d,e;z)
&=\frac{\Gamma(d)\Gamma(e)}{\Gamma(a)\Gamma(b)\Gamma(d-a)\Gamma(e-b)}
\int_0^1\int_0^1{\rm d}t_1{\rm d}t_2\notag \\
&\quad \times t_1^{a-1}t_2^{b-1}(1-t_1)^{d-a-1}(1-t_2)^{e-b-1}(1-t_1t_2z)^{-c}
\label{e:3F2integral2}\\
&=\frac{\Gamma(e)}{\Gamma(c)\Gamma(e-c)}
\int_0^1{\rm d}t\, t^{c-1}(1-t)^{e-c-1}{}_2F_1(a,b;d;zt)
\label{e:3F2integral1},\\
{}_4F_3(a,b,c,d;e,f,g;z)&=\frac{\Gamma(g)}{\Gamma(g-d)}
\int_0^1{\rm d}t\,
t^{d-1}(1-t)^{g-d-1}{}_3F_2(a,b,c;e,f;zt).
\label{e:4F3integral}
\end{align}
Conditions for convergence of these integrals are also given in Chapter~7 
of \cite{Prudnikov}.

In the case when $z=1$, we get
\begin{equation}
{}_2F_1(a,b;c;1)=\frac{\Gamma(c)\Gamma(c-a-b)}{\Gamma(c-a)\Gamma(c-b)}.
\label{e:2F1unity}
\end{equation}

\section{Explicit results for one-dimensional series}
\label{a:results}
The finite sums in the Theorem and Corollary of \secref{s:one-d} may be
further simplified for small values of $k$. 
One may then use properties of the Clausen
function given in \appref{a:specfunc} to reduce the sums to a single
Clausen function.
Here, we collect the results for such series involving
the psi function:
\begin{align}
&\sum_{n=1}^\infty\frac{1}{n^2}[\gamma+\psi(n)]=\zeta(3)
\label{e:psisum1}\\
&\sum_{n=1}^\infty\frac{1}{n^2}[\gamma+\psi(1+n)]=2\zeta(3)
\label{e:psisum2}\\  
&\sum_{n=1}^\infty\frac{1}{n^2}[\gamma+\psi(2n)]=\frac{9}{4}\zeta(3)
\label{e:psisum3}\\
&\sum_{n=1}^\infty\frac{1}{n^2}[\gamma+\psi(1+2n)]=\frac{11}{4}\zeta(3)
\label{e:psisum4}\\ 
&\sum_{n=1}^\infty\frac{1}{n^2}[\gamma+\psi(3n)]=
\frac{14}{3}\zeta(3)-\frac{2\pi}{3}\Cl{2}\left(\frac{\pi}{3}\right)\\
&\sum_{n=1}^\infty\frac{1}{n^2}[\gamma+\psi(1+3n)]=
5\zeta(3)-\frac{2\pi}{3}\Cl{2}\left(\frac{\pi}{3}\right)\\
&\sum_{n=1}^\infty\frac{1}{n^2}[\gamma+\psi(4n)]=
\frac{65}{8}\zeta(3)-2\pi G\\
&\sum_{n=1}^\infty\frac{1}{n^2}[\gamma+\psi(1+4n)]=
\frac{67}{8}\zeta(3)-2\pi G\\
&\sum_{n=1}^\infty\frac{1}{n^2}[\gamma+\psi(6n)]=
\frac{217}{12}\zeta(3)-\frac{16\pi}{3}\Cl{2}\left(\frac{\pi}{3}\right)\\
&\sum_{n=1}^\infty\frac{1}{n^2}[\gamma+\psi(1+6n)]=
\frac{73}{4}\zeta(3)-\frac{16\pi}{3}\Cl{2}\left(\frac{\pi}{3}\right)\\
&\sum_{n=1}^\infty\frac{(-1)^n}{n^2}[\gamma+\psi(n)]=\frac{1}{8}\zeta(3)
\label{e:altpsisum1}\\
&\sum_{n=1}^\infty\frac{(-1)^n}{n^2}[\gamma+\psi(1+n)]=-\frac{5}{8}\zeta(3)
\label{e:altpsisum2}\\
&\sum_{n=1}^\infty\frac{(-1)^n}{n^2}[\gamma+\psi(2n)]=
\frac{29}{16}\zeta(3)-\pi G\\
&\sum_{n=1}^\infty\frac{(-1)^n}{n^2}[\gamma+\psi(1+2n)]=
\frac{23}{16}\zeta(3)-\pi G\\ 
&\sum_{n=1}^\infty\frac{(-1)^n}{n^2}[\gamma+\psi(3n)]=
\frac{35}{8}\zeta(3)-2\pi\Cl{2}\left(\frac{\pi}{3}\right)\\
&\sum_{n=1}^\infty\frac{(-1)^n}{n^2}[\gamma+\psi(1+3n)]=
\frac{33}{8}\zeta(3)-2\pi\Cl{2}\left(\frac{\pi}{3}\right)
\end{align}
At least four of these results were probably known by Euler, in particular
the results of eqs.~\eqref{e:psisum1}, \eqref{e:psisum2},
\eqref{e:altpsisum1} and \eqref{e:altpsisum2}. A review of the history of 
this type of series, known as Euler series, can be found in \cite{Berndt}.

\section{The integral $I_p(x,a)$}
\label{a:integral}
In the proof of the theorem in \secref{s:one-d}, we will need a result 
for the integral \eqref{e:I-integral} in the limit $x\to 1^-$, which is
\begin{equation}
I_p(x,a)=\int_0^1{\rm d}t\, \frac{\Li{p}(a t)}{1-xt}
\end{equation}
for $p=2$. This integral is encountered for
arbitrary positive integers $p$ when studying the generalization of our 
theorem in \appref{a:generalization}. This section will be devoted to the 
study of this integral in the appropriate limit $x\to 1^-$. 
Integrating by parts, we obtain
\begin{equation*}
I_p(x,a)=-\frac{1}{x}\log(1-x)\Li{p}(a)
+\frac{1}{x}\int_0^1\frac{{\rm d}t}{t}\, \log(1-xt)\Li{p-1}(at).
\end{equation*}
Here, the singular part (as $x\to 1^-$) has been isolated, and we may
set $x=1$ in the remaining integral to get
\begin{equation*}
I_p(x,a)=-\frac{1}{x}\log(1-x)\Li{p}(a)
-\Lambda_{p-1}(a),
\end{equation*}
where 
\begin{equation}
\Lambda_{\nu}(a)=-\int_0^1\frac{{\rm d}t}{t}\, \log(1-t)\Li{\nu}(at).
\end{equation}
Here, we will need the following result
\begin{equation}
\Lambda_\nu(a)=S_{\nu,2}(a)+\Li{\nu+2}(a),
\label{e:Lambdaintegral}
\end{equation}
which can be proved by induction. By noting that $\Li{0}(z)=z/(1-z)$,
we find from eqs.~(3.12.7) and (2.2.5) of \cite{D&D} that the result holds 
for $\nu=0$.
Now, suppose the result is valid for $\nu-1$. 
Consider
\begin{equation*}
\Lambda_\nu(a)
=-\int_0^1\frac{{\rm d}t}{t}\, \log(1-t)\Li{\nu}(at)
=-\int_0^1\frac{{\rm d}t}{t}\, \log(1-t)
\int_0^1\frac{{\rm d}y}{y}\Li{\nu-1}(ayt),
\end{equation*}
which follows by eq.~(2.1.7) of \cite{D&D}.
By interchanging the order of integration
and invoking our assumption, we may perform the $t$ integration to get
\begin{equation*}
\Lambda_\nu(a)=\int_0^1\frac{{\rm d}y}{y}
\left[S_{\nu-1,2}(ay)+\Li{\nu+1}(ay)\right]
=S_{\nu,2}(a)+\Li{\nu+2}(a)
\end{equation*}
by once again using eq.~(2.1.7) of \cite{D&D}. 
By induction, the result \eqref{e:Lambdaintegral} holds for all non-negative 
integers $\nu$.

Thus, we conclude that
\begin{equation}
I_p(x,a)=-\frac{1}{x}\log(1-x)\Li{p}(a)-S_{p-1,2}(a)-\Li{p+1}(a)
\label{e:Iintegral}
\end{equation}
in the limit $x\to 1^-$.

\section{Generalizations}
\label{a:generalization}
One may foresee the need for generalizations of 
the results of the theorem, (\ref{e:singlepsisum}) and 
(\ref{e:singlepsisumalternating}), to powers $n^{-p}$ instead
of $n^{-2}$. Let us therefore consider the sum
\begin{equation}
\sigma_p(k)=\sum_{n=1}^\infty\frac{1}{n^p}\left[\gamma+\psi(1+kn)\right]
=\sum_{n=1}^\infty\sum_{j=1}^{kn}\frac{1}{n^pj},
\end{equation}
where $p=2,3,4,\dotsc$ and $k=1,2,3,\dotsc$.
Introducing the integral representation for the psi function,
and the regulator like in \secref{s:one-d}, we find
\begin{equation*}
\sigma_p(k)=\sum_{n=1}^\infty\frac{1}{n^p}
\int_0^1{\rm d}t\, \frac{1-t^{kn}}{1-t}
=\lim_{x\to 1^-}\int_0^1{\rm d}t\, \frac{\zeta(p)-\Li{p}(t^k)}{1-xt}.
\end{equation*}
For the polylogarithm, we use the factorization formula 
\eqref{e:polylogfactorization}
to get
\begin{equation*}
\sigma_p(k)=\lim_{x\to 1^-}\left[
-\frac{1}{x}\log(1-x)\zeta(p)-k^{p-1}\sum_{j=1}^k I_p(x,\omega^j)\right],
\label{e:Sp-split}
\end{equation*}
where we have introduced the integral
\begin{equation*}
I_p(x,a)=\int_0^1{\rm d}t\, \frac{\Li{p}(at)}{1-xt}. 
\end{equation*}
This integral was calculated in \appref{a:integral}.
Using this result and invoking the factorization
formula \eqref{e:polylogfactorization}, 
we note that the singular parts cancel. Thus, the result is
\begin{equation*}
\begin{split}
\sum_{n=1}^\infty\frac{1}{n^p}\left[\gamma+\psi(1+kn)\right]
&=k^{p-1}\sum_{j=1}^k\left[S_{p-1,2}(\omega^j)+\Li{p+1}(\omega^j)\right]\\
&=\frac{1}{k}\zeta(p+1)
+k^{p-1}\sum_{j=1}^k\, S_{p-1,2}(\omega^j).
\end{split}
\end{equation*}

As in the case with $p=2$, it is only the real part of the Nielsen
function 
that enters,
\begin{equation*}
S_{m,2}(\omega^j)+S_{m,2}(\omega^{-j}).
\end{equation*}
We have not been able to find any simple expression for this real
part for $m\ge2$. Thus, we leave the result as
\begin{equation}
\sum_{n=1}^\infty\frac{1}{n^p}\left[\gamma+\psi(1+kn)\right]
=\frac{1}{k}\, \zeta(p+1)
+k^{p-1}\sum_{j=1}^k\,  S_{p-1,2}(\omega^j).
\label{e:gen-singlepsisum}
\end{equation}

Similarly, for the generalization of the alternating series
\eqref{e:singlepsisumalternating}, we find
\begin{equation}
\sum_{n=1}^\infty\frac{(-1)^n}{n^p}\left[\gamma+\psi(1+kn)\right] 
=\frac{1}{k}\left(\frac{1}{2^{p}}-1\right) \zeta(p+1)
+k^{p-1}\sum_{j=1}^k\, S_{p-1,2}\left(\omega^{j+1/2}\right).
\label{e:altseriessum}
\end{equation}

When $p=1$, only the alternating series converges. We have been able 
to prove the following result,
\begin{equation}
\sum_{n=1}^\infty\frac{(-1)^n}{n}\left[\gamma+\psi(1+kn)\right]
=-\frac{1}{4}\left(k+\frac{1}{k}\right)\zeta(2)
+\frac{1}{2}\sum_{j=1}^k
\left\{\Cl{1}\left(\frac{2\pi j}{k}+\frac{\pi}{k}\right)\right\}^2,
\label{e:altseries1}
\end{equation}
and the immediate corollary thereof,
\begin{equation}
\sum_{n=1}^\infty\frac{(-1)^n}{n}\left[\gamma+\psi(kn)\right]
=\frac{1}{4}\left(\frac{1}{k}-k\right)\zeta(2)
+\frac{1}{2}\sum_{j=1}^k
\left\{\Cl{1}\left(\frac{2\pi j}{k}+\frac{\pi}{k}\right)\right\}^2.
\label{e:altseries2}
\end{equation}
\begin{proof}
By applying \eqref{e:altseriessum} we find that 
\begin{align*}
& \sum_{n=1}^\infty\frac{(-1)^n}{n}\left[\gamma+\psi(1+kn)\right]
=-\frac{1}{2k}\zeta(2)+\sum_{j=0}^{k-1}S_{0,2}(\omega^{j+1/2})\\
& =-\frac{1}{2k}\zeta(2)
+\frac{1}{2}\sum_{j=0}^{k-1}\left\{\Li{1}(\omega^{j+1/2})\right\}^2
\end{align*}
by making use of the fact that 
$S_{0,2}(z)=\tfrac{1}{2}\left\{\Li{1}(z)\right\}^2$. 
On the unit circle, we know from \cite{Lewin} that 
$\Li{1}(\theta)=\Cl{1}(\theta)+i\Gl{1}(\theta)$.
By retaining only the real parts of the expression, we arrive at
\begin{equation*}
-\frac{1}{2k}\zeta(2)
+\frac{1}{2}\sum_{j=0}^{k-1}
\left\{\left[\Cl{1}\left(\frac{2\pi j}{k}+\frac{\pi}{k}\right)\right]^2
-\left[\Gl{1}\left(\frac{2\pi j}{k}+\frac{\pi}{k}\right)\right]^2\right\}.
\end{equation*}
The function $\Gl{1}(\theta)$ is a well-known Fourier series equal to the 
$2\pi$-periodic extension of $\tfrac{1}{2}(\pi-\theta),\ 
0<\theta<2\pi$. This fact is used to perform the finite sum
over the part containing $\Gl{1}(\theta)$,
\begin{align*} 
&-\frac{1}{2k}\zeta(2)
-\frac{\pi^2}{8}\sum_{j=0}^{k-1}\left(1-\frac{2j}{k}-\frac{1}{k}\right)^2
+\frac{1}{2}\sum_{j=0}^{k-1}
\left\{\Cl{1}\left(\frac{2\pi j}{k}+\frac{\pi}{k}\right)\right\}^2\\
&=-\frac{1}{4}\left(k+\frac{1}{k}\right)\zeta(2)
+\frac{1}{2}\sum_{j=1}^k
\left\{\Cl{1}\left(\frac{2\pi j}{k}+\frac{\pi}{k}\right)\right\}^2.
\end{align*}
\end{proof}
By using the fact that $\Cl{1}(\theta)=-\log\abs{2\sin{\tfrac{\theta}{2}}}$, 
we may now state the following results:
\begin{align}
&\sum_{n=1}^\infty\frac{(-1)^n}{n}\left[\gamma+\psi(n)\right]=
\frac{1}{2}\log^2 2\\
&\sum_{n=1}^\infty\frac{(-1)^n}{n}\left[\gamma+\psi(1+n)\right]=
-\frac{1}{2}\zeta(2)+\frac{1}{2}\log^2 2\\
&\sum_{n=1}^\infty\frac{(-1)^n}{n}\left[\gamma+\psi(2n)\right]=
-\frac{3}{8}\zeta(2)+\frac{1}{4}\log^2 2\\
&\sum_{n=1}^\infty\frac{(-1)^n}{n}\left[\gamma+\psi(1+2n)\right]=
-\frac{5}{8}\zeta(2)+\frac{1}{4}\log^2 2\\
&\sum_{n=1}^\infty\frac{(-1)^n}{n}\left[\gamma+\psi(3n)\right]=
-\frac{2}{3}\zeta(2)+\frac{1}{2}\log^2 2\\
&\sum_{n=1}^\infty\frac{(-1)^n}{n}\left[\gamma+\psi(1+3n)\right]=
-\frac{5}{6}\zeta(2)+\frac{1}{2}\log^2 2
\end{align}
For higher values of $k$, the sums do not turn out to be this simple.


\end{document}